\documentclass[a4paper, 11pt]{article}
\pdfoutput=1

\usepackage{jheppub}
\usepackage[utf8]{inputenc}
\usepackage[T1]{fontenc}

\newcommand{\Dmq}{\Delta m^2}
\newcommand{\Nuc}[2]{{\ensuremath{\mbox{}^{#1}}\text{#2}}}
\newcommand{\eVq}{\ensuremath{\text{eV}^2}}

\makeatletter
\DeclareRobustCommand\recite[1]{\begingroup\@fileswfalse\cite{#1}\endgroup}
\makeatother

\graphicspath{{Figures/}}

\allowdisplaybreaks

\title{Lessons from the first JUNO results}

\author[a,b]{Ivan Esteban,}
\affiliation[a]{Department of Physics, University of the Basque
  Country UPV/EHU, PO Box 644, 48080 Bilbao, Spain}
\affiliation[b]{EHU Quantum Center, University of the Basque Country
  UPV/EHU}
\emailAdd{ivan.esteban@ehu.eus}

\author[c,d,e]{M.~C.~Gonzalez-Garcia,}
\affiliation[c]{Departament de F\'{\i}sica Qu\`antica i
  Astrof\'{\i}sica and Institut de Ciencies del Cosmos, Universitat de
  Barcelona, Diagonal 647, E-08028 Barcelona, Spain}
\affiliation[d]{Instituci\'o Catalana de Recerca i Estudis
  Avan\c{c}ats (ICREA), Pg.\ Lluis Companys 23, 08010 Barcelona,
  Spain.}
\affiliation[e]{C.N.~Yang Institute for Theoretical Physics, State
  University of New York at Stony Brook, Stony Brook, NY 11794-3840,
  USA}
\emailAdd{concha.gonzalez-garcia@stonybrook.edu}

\author[f]{Michele Maltoni,}
\affiliation[f]{Instituto de F\'isica Te\'orica (IFT-CFTMAT), CSIC-UAM,
  Calle de Nicol\'as Cabrera 13--15, Campus de Cantoblanco, E-28049
  Madrid, Spain}
\emailAdd{michele.maltoni@csic.es}

\author[g]{Ivan Martinez-Soler,}
\affiliation[g]{Institute for Particle Physics Phenomenology, Durham
  University, South Road, DH1 3LE, Durham, UK}
\emailAdd{ivan.j.martinez-soler@durham.ac.uk}

\author[h,i]{Jo\~{a}o Paulo Pinheiro,}
\affiliation[h]{State Key Laboratory of Dark Matter Physics, Tsung-Dao
  Lee Institute \& School of Physics and Astronomy, Shanghai Jiao Tong
  University, Shanghai 200240, China}
\affiliation[i]{Key Laboratory for Particle Astrophysics and Cosmology
  (MOE) \& Shanghai Key Laboratory for Particle Physics and Cosmology,
  Shanghai Jiao Tong University, Shanghai 200240, China}
\emailAdd{joaopaulo.pinheiro@fqa.ub.edu}

\author[j]{Thomas Schwetz}
\affiliation[j]{Institut f\"ur Astroteilchenphysik, Karlsruher
  Institut für Technologie (KIT), Hermann-von-Helmholtz-Platz 1, 76344 Eggenstein-Leopoldshafen, Germany}
\emailAdd{schwetz@kit.edu}

\abstract{First results from the JUNO reactor neutrino experiment
  already determine with world-leading precision the small neutrino
  squared-mass splitting $\Dmq_{21}$ and the mixing angle
  $\theta_{12}$.  In this article we perform an exploratory study
  beyond these, taking advantage of the first JUNO data release to
  discuss its sensitivity to the large squared-mass splitting,
  $\Dmq_{3\ell}$.  When combined with constraints from global
  oscillation data, this may already contain some information on the
  neutrino mass ordering.  Indeed, we find that the combination of the
  complementary $\Dmq_{3\ell}$-determinations gives a slight
  preference for Normal Ordering, with a $p$-value for Inverted
  Ordering of 2\%--2.6\% ($2.2\sigma$--$2.3\sigma$).  We study the
  robustness of this result with respect to potential systematic
  uncertainties and statistical fluctuations.  Taken at face value, a
  full global analysis of oscillation data including the publicly
  available JUNO information and data leads to a preference for Normal
  Ordering with $\Delta\chi^2 = 4.6$ and 9.4 without and with Super-K
  and IceCube-24 atmospheric neutrino data, respectively.}

\preprint{IFT-UAM/CSIC-26-3, IPPP/26/03, YITP-SB-2026-02}


\begin{document}

\maketitle

\section{Introduction}

In the realm of neutrino oscillation studies, the first results of the
Jiangmen Underground Neutrino Observatory (JUNO)~\cite{JUNO:2025gmd}
represent a significant step forward.  Just 59.1 days of exposure have
been enough to reach world-leading precision on their dominant
oscillation parameters, $\Dmq_{21}$ and $\theta_{12}$.  These first
results constitute a milestone in the path to achieve JUNO's ultimate
goals of precision measurements of $\Dmq_{21}$, $\theta_{12}$, and
$\Dmq_{31}$~\cite{JUNO:2022mxj}, and of determining the neutrino mass
ordering (MO) with more than $3\sigma$
significance~\cite{JUNO:2015zny}.

Determining the MO is an important open question in neutrino physics.
Among others, it has significant implications for neutrinoless double
beta decay searches, it can affect the neutrino mass scale relevant
for cosmology, and it is key to break existing degeneracies,
\textit{e.g.}, in leptonic CP violation~\cite{Esteban:2024eli}.  While
the intrinsic sensitivity of JUNO to the MO is based on a subtle
interference effect between fast ($\Dmq_{31}$-driven) and slow
($\Dmq_{21}$-driven) oscillation modes~\cite{Petcov:2001sy,
  Choubey:2003qx}, a potentially much earlier identification of the MO
could emerge from the combination of independent percent-level
determinations of $|\Dmq_{31}|$ from JUNO and $\nu_\mu/\bar{\nu}_\mu$
disappearance data.  This latter possibility was mentioned in
Refs.~\cite{Choubey:2003qx, deGouvea:2005hk} and demonstrated in
detail in Ref.~\cite{Nunokawa:2005nx}, see also~\cite{Minakata:2006gq,
  Blennow:2013vta, Cabrera:2020ksc,Parke:2024xre}.

In this respect, it is important to remark that in
Ref.~\cite{JUNO:2025gmd} the JUNO collaboration has not presented any
results on $|\Dmq_{31}|$ and the MO with their first data, limiting
the analysis (including the provided details on the systematic
uncertainties) to the dominant oscillation mode due to $\Dmq_{21}$ and
$\theta_{12}$.  Nevertheless, it can be an informative exercise to
explore the possible sensitivity to $|\Dmq_{31}|$ and the MO of these
first results within the publicly available information, as well as
the dependence of these results on possible unknowns, especially in
combination with the remaining global oscillation data.

With these goals in mind, in this paper we present our independent
implementation of the analysis of the first JUNO data, which we
describe in Sec.~\ref{sec:ana}.  We perform a variety of analyses in
order to reproduce the results of the collaboration on $\Dmq_{21}$ and
$\theta_{12}$, which have been incorporated in our updated global
analysis NuFIT-6.1~\cite{nufit-6.1, Esteban:2024eli}
(Sec.~\ref{sec:solar}).  With this well-tuned tool in hand, we make an
exploratory study of the possible sensitivity to the large
squared-mass splitting $\Dmq_{3\ell}$, discussing both its absolute
value as well as the MO (Sec.~\ref{sec:order}).  We study its
robustness against statistical fluctuations by running a Monte Carlo
simulation of JUNO data in Sec.~\ref{sec:MC}, which allows us to
perform a statistically consistent MO hypothesis test.  We further
investigate the potential impact of various systematic uncertainties
in Sec.~\ref{sec:robust}.  In the summary, Sec.~\ref{sec:summary}, we
also comment on the MO discrimination of the global oscillation data
including our attempted JUNO analysis.

In the following, we adopt the standard parametrization of the
$3\times 3$ unitary leptonic mixing matrix~\cite{Maki:1962mu,
  Kobayashi:1973fv}, as discussed in Ref.~\cite{Esteban:2024eli}.  The
$\bar\nu_e$ survival probability relevant for reactor experiments such
as JUNO depends on the four parameters $\{\theta_{12}, \theta_{13},
\Dmq_{21}, \Dmq_{3\ell}\}$, with $\Dmq_{ij} \equiv m^2_j - m^2_i$ and
$m_{1,2,3}$ the masses of the three neutrino mass eigenstates.  There
are two non-equivalent orderings for the three neutrino masses: normal
ordering (NO) with $m_1 < m_2 < m_3$, and inverted ordering (IO) with
$m_3 < m_1 < m_2$.  Thus one can parametrize the ordering in terms of
the sign of $\Dmq_{3\ell}$ defined as
\begin{equation}
  \Dmq_{3\ell}
  \quad \text{with}\quad
  \begin{cases}
    \ell = 1 & \text{for $\Dmq_{3\ell} > 0$: normal ordering (NO),} \\
    \ell = 2 & \text{for $\Dmq_{3\ell} < 0$: inverted ordering (IO).}
  \end{cases}
\end{equation}
Recent global analyses within the three-flavour scenario can be found
in Refs.~\cite{Esteban:2024eli, nufit-6.1, Capozzi:2025wyn,
  deSalas:2020pgw}.


\section{JUNO analysis details}
\label{sec:ana}

In this Section, we describe our analysis of JUNO data that allows us
to obtain the main quantitative results of this work.  The neutrino
detection reaction in JUNO is inverse beta decay (IBD), $\bar\nu_e + p
\to e^+ + n$, and the energy deposited by the positron (so-called
``prompt energy'', see below) allows to infer the neutrino energy.  We
fit the 66 data points in reconstructed prompt energy provided in
Fig.~3 of Ref.~\cite{JUNO:2025gmd}, 2379 events in total,
corresponding to 59.1 days of data taking.  The basic ingredients for
the analysis are as follows.

\paragraph{Signal spectrum prediction.}
We compute the expected number of IBD signal events $N^S_i$ in the
reconstructed prompt energy bin $i$ as a function of oscillation
parameters $\vec\omega$:
\begin{equation}
  \label{eq:signal}
  N^S_i(\vec\omega)= C\,
  \sum_r \int dE_\nu  \frac{\mathcal{P}_r}{4 \pi L_r^2}\,
  \phi(E_\nu)\, \sigma_\textsc{ibd}(E_\nu)\,
  P_{ee}(E_\nu,L_r,\vec\omega)\, R_i(E_\nu) \,,
\end{equation}
where the sum in $r$ extends over the nine reactors given in Tab.~2
of~\cite{JUNO:2022mxj} and $L_r$ is the distance to each of them.
This includes eight reactors at a distance of about 53~km and a single
effective reactor from the Daya Bay complex at 215~km.  The
contribution of each reactor is weighted by its average power
$\mathcal{P}_r$ as given in the same table.  In the absence of
information on the relevant isotope composition contributing to the
neutrino flux for the present JUNO data set, we adopt the same as for
Daya Bay for all reactors.  For the emitted flux per unit power,
$\phi(E_\nu)$, we assume the un-oscillated flux extracted from data by
the Daya Bay collaboration, including 25 pulls to parametrise the
uncertainty, as described in Appendix~\ref{app:fluxreact}.  In
Eq.~\eqref{eq:signal} $C$ is a normalization constant accounting for
the number of target protons, lifetime and selection efficiencies;
$\sigma_\textsc{ibd}(E_\nu)$ is the IBD cross section that we obtain
from Ref.~\cite{Vogel:1999zy} (we have checked that using the cross
sections from Refs.~\cite{Strumia:2003zx, Tomalak:2025jtn} leads to
very similar results); $R_i(E_\nu)$ is the energy response function
for bin $i$, that we describe below; and $P_{ee}(E_\nu, \vec\omega)$
is the neutrino flavor oscillation probability.  For this last
quantity, we use the analytic expression for 3-flavour oscillations
\begin{equation}
  \label{eq:Pee}
  \begin{split}
    P_{ee} = 1
    &- c_{13}^4 \sin^22\theta_{12}\sin^2\frac{\Dmq_{21}L}{4E_\nu}
    \\
    &- \sin^22\theta_{13}\left(c_{12}^2 \sin^2\frac{\Dmq_{31}L}{4E_\nu}
    + s_{12}^2 \sin^2\frac{\Dmq_{32}L}{4E_\nu} \right) ,
  \end{split}
\end{equation}
with $s_{ij} \equiv \sin\theta_{ij}$ and $c_{ij} \equiv
\cos\theta_{ij}$.  This expression holds for both mass orderings.  We
include matter effects by introducing effective mixing parameters in
matter, expanding the dominant 12-term to linear order in the small
parameter $A = 2E_\nu V / \Dmq_{21}$.  We do not take into account
subleading matter effects on the fast-oscillating terms in the second
line of Eq.~\eqref{eq:Pee}~\cite{Capozzi:2013psa}.  The resulting
probability agrees within better than 0.4\% with a full numerical
calculation.  We assume a constant matter density of
$2.55~\text{g}/\text{cm}^3$~\cite{JUNO:2025gmd}, see also
Ref.~\cite{Li:2016txk, Khan:2019doq, Li:2025hye}. For all analyses
below (except the global analysis results in Sec.~\ref{sec:summary})
we assume $\sin^2\theta_{13} = 0.022 \pm
0.00056$~\cite{Esteban:2024eli}, taking into account the uncertainty
by introducing a Gaussian pull (the results are very similar if
$\theta_{13}$ is instead fixed).

Flux, cross section, and oscillation probability are calculated for a
true neutrino energy $E_\nu$.  The observable in the detector is the
energy deposited by the positron after annihilation, the so-called
``prompt energy'' $E_\text{pr} = E_e + m_e$ where $E_e$ is the
positron energy, that is related to $E_\nu$ by the kinematics of IBD
-- see, \textit{e.g.}, Ref.~\cite{Strumia:2003zx}.  This relationship,
together with the energy resolution of the detector, is encapsulated
in the function $R_i(E_\nu)$ in Eq.~\eqref{eq:signal}.

In the limit of neglecting the momentum carried away by the recoiling
neutron, there is a one-to-one relation between neutrino and prompt
energies, $E_\text{pr} = E_\nu + m_e - \Delta$ with $\Delta =
m_n-m_p$.  In this limit, $R_i(E_\nu)$ is given by a convolution of a
Dirac delta linking $E_\nu$ and $E_\text{pr}$ and the Gaussian energy
resolution of the detector.  However, as discussed in
Ref.~\cite{Capozzi:2013psa}, for precision reactor experiments this is
not a good approximation and neutron-recoil effects need to be taken
into account.  In our analysis, we include this by following the
prescription described in Ref.~\cite{Capozzi:2013psa}.  In detail, if
$E_e^\text{inf}(E_\nu)$ and $E_e^\text{sup}(E_\nu)$ are the minimum
and maximum kinematically allowed positron energies for a given
neutrino energy $E_\nu$ (see, \textit{e.g.},
Ref.~\cite{Strumia:2003zx} for explicit expressions), we obtain
$R_i(E_\nu)$ by convolving a normalized top-hat function between
$E_\text{pr}^\text{inf} = E_e^\text{inf} + m_e$ and
$E_\text{pr}^\text{sup} = E_e^\text{sup} + m_e$ with the Gaussian
energy resolution of the detector (see Eqs.~(35) and~(36) in
Ref.~\cite{Capozzi:2013psa} for explicit
expressions).\footnote{Alternatively, we have checked that the main
effect of the neutron recoil is a slight shift of $E_e$ from $E_\nu -
\Delta$, which can be approximately captured by setting $E_e(E_\nu) =
(E_e^\text{inf} + E_e^\text{sup})/2$.}

Regarding the energy resolution of the detector, following Eq.~(8) in
Ref.~\cite{JUNO:2025fpc}, we adopt a Gaussian resolution of width
\begin{equation}
  \label{eq:resol}
  \sigma(E_\text{pr}) = E_\text{pr} \sqrt{a^2/E_\text{pr} + b^2} \,,
\end{equation}
with $a=0.033$, $b=0.01$ for $E_\text{pr}$ in MeV.  We also account
for non-linearity in the detector energy response by modifying
$E_\text{pr}^\text{inf}(E_\nu)$ and $E_\text{pr}^\text{sup}(E_\nu)$ as
$E_\text{pr} \to E_\text{pr}\, F_\text{n.l.}(E_\text{pr})$, where
$F_\text{n.l.}(E_\text{pr})$ is the non-linearity function for
positrons, that we extract from Fig.~6d in Ref.~\cite{JUNO:2025gmd}
(see also Ref.~\cite{JUNO:2025fpc}).

\paragraph{Backgrounds.}
Following Ref.~\cite{JUNO:2025gmd}, we separate the background into
five components: $\Nuc{9}{Li} \big/ \Nuc{8}{He}$ produced by cosmic
muon spallation, that we denote as ``LiHe''; geoneutrinos, that we
denote as ``Geo''; world reactors, that we denote as ``world-reac'';
$\Nuc{214}{Bi} \big/ \Nuc{214}{Po}$ from radon decay, that we denote
as ``BiPo''; and other sources of background, that we denote as
``others''.  The number of events of each of these backgrounds in bin
$i$ ($N_i^\text{LiHe}$, $N_i^\text{Geo}$, $N_i^\text{world-reac}$,
$N_i^\text{BiPo}$, and $N_i^\text{others}$) is read from Fig.~3 of
Ref.~\cite{JUNO:2025gmd}, with the exception of world reactors for
which we take the un-oscillated reactor spectrum (since oscillations
are averaged out for very far reactors, so the oscillated shape is
proportional to the un-oscillated one).  We normalize all background
spectra to the total pre-fit rate for each background given in Tab.~1
of Ref.~\cite{JUNO:2025gmd}.

\begin{figure}
  \centering
  \includegraphics[width=0.49\textwidth]{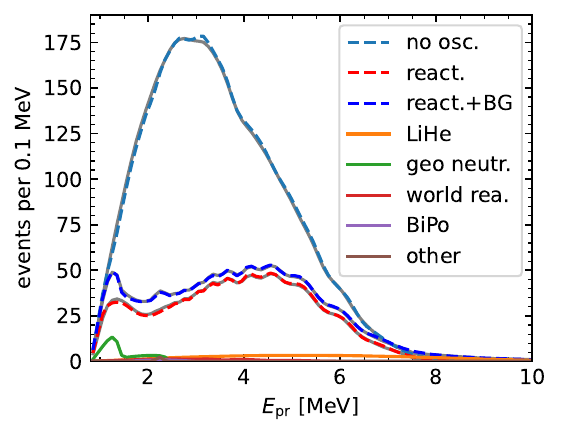}\hfill
  \includegraphics[width=0.49\textwidth]{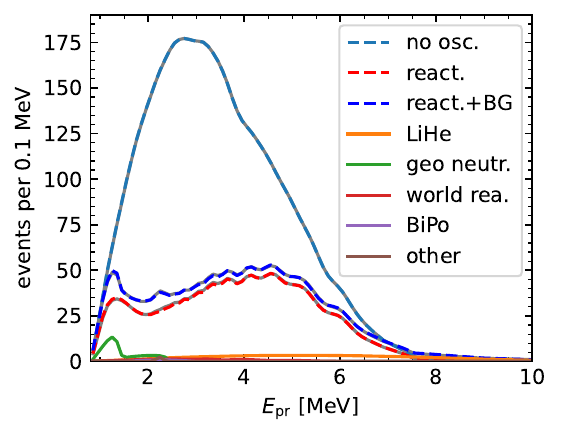}
  \caption{Predicted spectra (dashed) compared to the official JUNO
    ones~\cite{JUNO:2025gmd} (grey).  Left: based on our own predicted
    spectrum, normalized to match the normalization of their
    un-oscillated spectrum.  Right: bin-per-bin rescaled to match the
    JUNO un-oscillated spectrum.}
  \label{fig:spectrum}
\end{figure}

With all these elements, we obtain our predicted un-oscillated and
best-fit oscillated spectra shown in dashed lines in the left panel of
Fig.~\ref{fig:spectrum}.  In these predictions, the normalization
constant $C$ has been set to match the normalization of the JUNO
un-oscillated spectrum.  As seen in the figure, the predicted spectra
agree very well with the official JUNO results (shown in grey).  In
the right panel, we scale our predictions bin-per-bin to match the
JUNO un-oscillated spectrum, which gives a slightly better match to
their oscillation parameter regions.  Such minor tuning is inevitable
given the lack of publicly available information on, \textit{e.g.},
the exact isotope composition and power of the relevant nuclear
reactors.

\paragraph{Systematics.}
The following sources of systematic uncertainties $j$ are included as
pulls $\xi_j$ in the statistical analysis described below, assuming
Gaussian errors $\sigma_j$ as follows:
\begin{itemize}
\item Normalization of the reactor neutrino event rate:
  $\sigma_\text{norm} = 1.8\%$, obtained by summing in quadrature all
  entries in Tab.~2 of Ref.~\cite{JUNO:2025gmd}.

\item Normalization of each background component:
  $\sigma_{\text{bg},i} = 33\%$, $42\%$, $10\%$, $56\%$, $100\%$ for
  $i=$ LiHe, Geo, world-reac, BiPo, other, respectively, as given in
  Tab.~1 of Ref.~\cite{JUNO:2025gmd}.

\item Shape uncertainty for the LiHe background: set to 20\% at 1~MeV
  and linearly proportional to energy~\cite{JUNO:2025gmd}.

\item 25 pulls for the reactor flux uncertainty, as explained in
  Appendix~\ref{app:fluxreact}.

\item Energy scale uncertainties: we introduce two pulls
  $\xi_\text{scl}$ and $\xi_\text{bias}$, which shift prompt energy as
  \begin{equation}
    \label{eq:Eshift}
    E_\text{pr} \to \tilde{E}_\text{pr} \equiv  E_\text{pr}
    \big[ (1+\xi_\text{scl})\, F_\text{n.l.}(E_\text{pr})
      + \xi_\text{bias} \big]
  \end{equation}
  where $\xi_\text{scl}$ and $\xi_\text{bias}$ parametrize an
  uncertainty in the energy scale and an additive bias in the
  non-linearity correction $F_\text{n.l.}$, respectively.  We find
  that the effect of both pulls is very similar when their uncertainty
  is kept at the nominal value $\sigma_\text{scl} = \sigma_\text{bias}
  = 0.5\%$~\cite{JUNO:2025gmd}.  Therefore, in most of our analysis
  cases we only use one of them at a time, typically $\xi_\text{scl}$
  (we discuss this further in Sec.~\ref{sec:robust}).

\item Energy resolution uncertainty: we introduce an uncertainty on
  the energy resolution by replacing $\sigma \to \tilde\sigma \equiv
  (1 + \xi_\text{res}) \sigma$ in the response function $R_i(E_\nu)$,
  where $\sigma$ is given in Eq.~\eqref{eq:resol}.  As we have not
  found publicly available information on the JUNO energy resolution
  uncertainty, we adopt a nominal uncertainty $\sigma_\text{res} =
  5\%$.  In Sec.~\ref{sec:robust}, we assess the impact of varying
  this uncertainty.
\end{itemize}
In addition, in some of the analysis variants we introduce fixed
``ad-hoc'' rescaling factors ${r}_i$ for some of the backgrounds, for
$F_\text{n.l.}$, and for the energy resolution $\sigma$ in
Eq.~\eqref{eq:resol}.  Altogether, our predicted number of signal and
background events in presence of the systematic pulls and rescaling
factors is
\begin{equation}
  \label{eq:Nshift}
  \begin{aligned}
    N_i^S(\vec\omega, \vec\xi, \vec{r})
    &= (1 + \xi_\text{norm})\, \tilde{N}_i^S (\vec\omega, \xi_\text{scl},
    \xi_\text{bias}, \xi_\text{res}, r_\text{n.l.}, r_\text{res}) \,,
    \\
    N^\text{Geo}_i(\vec\xi, \vec{r})
    &= (1+\xi_\text{Geo})\, N^\text{Geo}_i \,,
    \\
    N^\text{LiHe}_i(\vec\xi, \vec{r})
    &= (1 + \xi_\text{LiHe,1} + \xi_\text{LiHe,2}\, E_i)\,
    r_\text{BG}\, N^\text{LiHe}_i \,,
    \\
    N^\text{BiPo}_i(\vec\xi, \vec{r})
    &= (1 + \xi_\text{BiPo})\, r_\text{BG}\, N^\text{BiPo}_i
    \\
    N^\text{world-reac}_i(\vec\xi, \vec{r})
    &= (1 + \xi_\text{world-reac})\, r_\text{BG}\, N^\text{world-reac}_i \,,
    \\
    N^\text{other}_i(\vec\xi, \vec{r})
    &= (1 + \xi_\text{other})\, r_\text{BG}\, N^\text{other}_i
  \end{aligned}
\end{equation}
where $\tilde{N}_i^S$ in the right hand side of the first line is
obtained as in Eq.~\eqref{eq:signal} with modified reconstructed
energy $\tilde{E}_\text{pr}$ and energy resolution $\tilde\sigma$ as
\begin{align}
  \label{eq:rnl}
  \tilde{E}_\text{pr} &= E_\text{pr} \, r_\text{n.l.}
  \big[ (1 + \xi_\text{scl})\, F_\text{n.l.}(E_\text{pr})
    + \xi_\text{bias} \big] \,,
  \\
  \label{eq:rres}
  \tilde\sigma &= (1 + \xi_\text{res})\, r_\text{res}\, \sigma\,.
\end{align}

\paragraph{Statistical analysis.}
With all these elements we build our $\chi^2$ function for the
oscillation parameters ($\vec\omega$), where we include the pull
uncertainties as Gaussian penalties
\begin{equation}
  \chi^2_\text{JUNO}(\vec\omega)
  = \min_{\vec\xi} \Bigg[ \chi^2_\text{data}(\vec\omega, \vec\xi)
    + \sum_j \frac{\xi_j^2}{\sigma_{\xi_j}^2} \Bigg] \,.
\end{equation}
In Ref.~\cite{JUNO:2025gmd} the JUNO collaboration adopted the
so-called CNP definition~\cite{Ji:2019yca} for
$\chi^2_\text{data}(\vec\omega, \xi_i)$, defined by $\chi^2_\text{CNP}
\equiv \sum_i (P_i - O_i)^2 \big/ \sigma_i^2$ where $O_i$ and $P_i$
are the observed and predicted number of events in bin $i$,
respectively, and
\begin{equation}
  \sigma^2_i = \frac{3}{\frac{1}{O_i} + \frac{2}{P_i}} \,.
\end{equation}
The CNP $\chi^2$ has been constructed in order to approximate the
Poisson $\chi^2$, defined by $\chi^2_\text{Poisson} \equiv 2 \sum_i
[P_i - O_i + O_i \log(O_i/P_i)]$.  We find that, while
$\chi^2_\text{CNP}$ gives results very similar to
$\chi^2_\text{Poisson}$, they are not identical.  Below, we show
results for different analyses based on either of these two
definitions.


\section{Determination of $\Dmq_{21}$ and $\theta_{12}$}
\label{sec:solar}

\begin{table}
  \small\centering
  \begin{tabular}{|c|cc|cccc|}
    \hline
    & cnf~1 & cnf~2 & cnf~3 & cnf~4 & cnf~5 & cnf~6 \\
    \hline
    $r_\text{BG}$ & 1 & 1.15 & 1.15 & 1.15 & 1.& 1.15 \\
    $r_\text{n.l.}$ & 1 & 1 & 1.024 & 1 & 1 & 1 \\
    $\sigma_\text{bias}$ (\%) & 0 & 0 & 0 & 5 & 0 & 0 \\
    $r_\text{res}$ & 1 & 1 &1 &1 &1.3 & 1 \\
    $\sigma_\text{res}$ (\%) & 5 & 5 & 5 & 5 & 5 & 40
    \\
    $\chi^2_\text{data}$ & CNP
    & Poisson & Poisson & Poisson & Poisson & Poisson
    \\
    \hline
    $\chi^2_\text{min}$ & 49.4 & 49.2 & 48.9 & 49.2 & 50.1 & 49.1
    \\
    \hline\hline
    $\Delta\chi^2_\mathrm{IO-NO}$ & 3.18 (3.41) & 3.05 (3.28)
    & 0.30 (1.60) & 2.89 (3.03) & 2.08 (2.21) & 2.06 (2.11)
    \\
    $T_0^\text{NO}$ & 1.37 (1.72) & 1.35 (1.69)
    & 1.40 (1.70) & 1.30 (1.65) & 0.69 (0.84) & 1.29 (1.58)
    \\
    $-T_0^\text{IO}$ & 1.45 (1.69) & 1.43 (1.66)
    & 1.46 (1.65) & 1.34 (1.56) & 0.71 (0.82) & 1.34 (1.53)
    \\
    \hline
  \end{tabular}
  \caption{Summary of different analysis configurations cnf~1--6, see
    Sec.~\ref{sec:ana} for definitions.  Our default configuration is
    cnf~2.  For all configurations, $\sigma_\text{norm} = 1.8\%$ and
    $\sigma_\text{scl} = 0.5\%$.  $r_\text{BG}$ is a rescaling of all
    backgrounds except for geoneutrinos.  $\chi^2_{\min}$ is given for
    JUNO only; the dof are 63 (66 bins minus 3 fitted parameters).
    $\Delta\chi^2_\mathrm{IO-NO}$ and $T_0$ values are for JUNO
    combined with the $|\Dmq_{3\ell}|$ determination from NuFIT-6.1
    without (with) SK-ATM.  $T_0$ corresponds to
    $\Delta\chi^2_\mathrm{IO-NO}$ for the Asimov data set assuming
    NuFIT-6.1 best-fit values for $\Dmq_{3\ell}$; see
    Sec.~\ref{sec:MC} for a translation to $p$-values.}
  \label{tab:conf}
\end{table}

Having specified the details of our JUNO data analysis, we first apply
it to determining $\Dmq_{21}$ and $\theta_{12}$, comparing our results
with the official ones from the collaboration.  We have performed
several analysis variants with different configurations for some of
the variables discussed above.  In this paper, we present the
representative results of six of them, denoted as cnf~1 to cnf~6.  We
list in Table~\ref{tab:conf} the different assumptions about the
parameters changed in the fit.

\begin{figure}
  \begin{minipage}[t]{0.58\textwidth}
    \vspace{0pt}
    \centering
    \includegraphics[width=\linewidth]{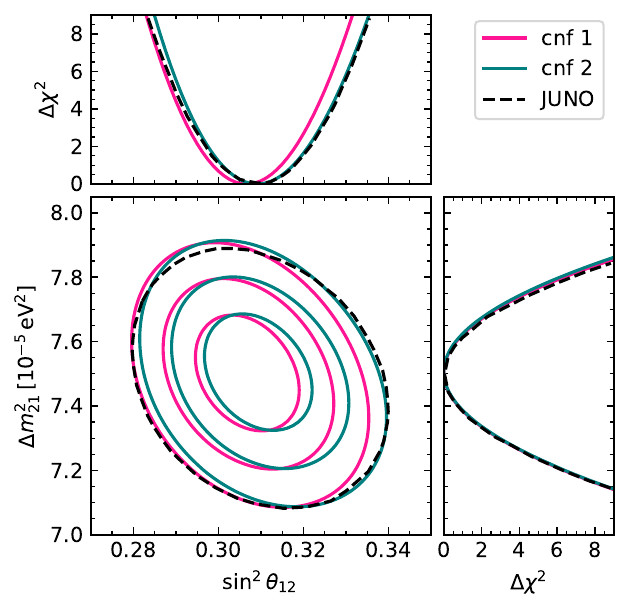}
    \\[-82mm]
  \end{minipage}
  \hfill
  \begin{minipage}[t]{0.38\textwidth}
    \vspace{0pt}
    \centering
    \includegraphics[width=\linewidth]{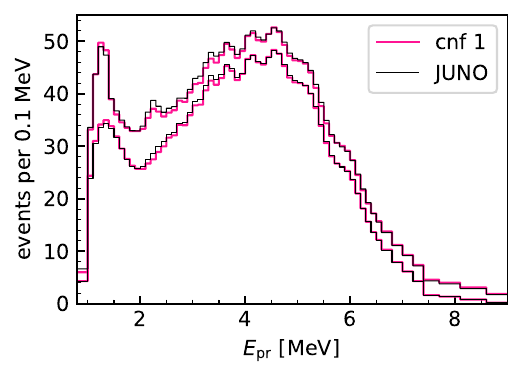}\\
    \includegraphics[width=\linewidth]{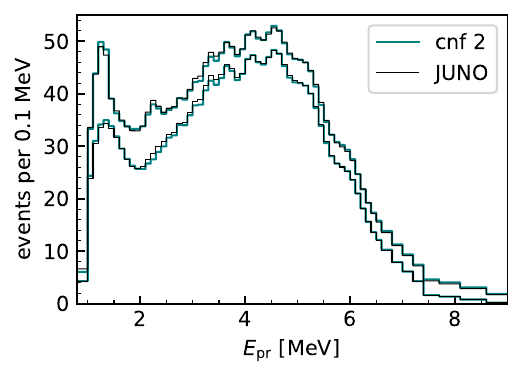}
  \end{minipage}
  \caption{Left: Determination of $\Dmq_{21}$ and $\sin^2\theta_{12}$
    for the two configurations cnf~1 and cnf~2, compared to the JUNO
    results (black dashed line).  Contours are for $1\sigma$,
    $2\sigma$, and $3\sigma$ (2 dof).  Right: Best-fit reactor
    neutrino spectra without pull shifts (lower histograms) and
    reactor neutrino+background spectra with pull shifts (higher
    histograms) for cnf~1 (upper) and cnf~2 (lower).}
  \label{fig:par12-1-2}
\end{figure}

The first two variants, cnf~1 and cnf~2, illustrate our choices to
reproduce the official $\Dmq_{21}$ and $\theta_{12}$ determination.
We show the results for these configurations in
Fig.~\ref{fig:par12-1-2}.  For each configuration, we compare with the
official JUNO results for the determination of $\Dmq_{21}$ and
$\sin^2\theta_{12}$ as well as for the predicted spectra at the
best-fit parameters quoted in Ref.~\cite{JUNO:2025gmd}.  For the
spectra shown in the right panels, we show the predicted reactor
spectrum without pull shifts and the reactor+background spectrum
including the pull shifts.  In all cases, we have performed the
analysis both by fixing $\Dmq_{3\ell}$ and by marginalizing over it,
and we have verified that the ($\Dmq_{21}$, $\sin^2\theta_{12}$)
regions are unchanged within the precision of the plot, in agreement
with Ref.~\cite{JUNO:2025gmd} (see also a corresponding discussion in
Ref.~\cite{Capozzi:2025ovi}).

The first configuration, cnf~1, corresponds to all pull uncertainties
at their nominal values and no additional rescaling factors (after
rescaling bin-by-bin the un-oscillated spectrum).  Following the JUNO
collaboration, we also adopt the CNP definition of $\chi^2$.  From
Fig.~\ref{fig:par12-1-2}, we see that the allowed regions in
($\Dmq_{21}$, $\sin^2\theta_{12}$) are close to the ones of JUNO with
a slightly smaller $\sin^2\theta_{12}$.  In order to compensate for
this, we have increased all backgrounds except for geoneutrinos by
15\%.  After additionally changing the $\chi^2$ definition from CNP to
Poisson, this corresponds to the configuration cnf~2.  We see that,
with this minor adjustment, cnf~2 provides an excellent reproduction
of the official JUNO results.  This background rescaling is well
within their $1\sigma$ uncertainties, and it also leads to a slightly
better reproduction of the best-fit JUNO spectrum shown in the right
panels of Fig.~\ref{fig:par12-1-2}, providing additional justification
for increasing them.  We adopt cnf~2 as our default configuration for
further explorations.  We notice in passing that all versions give a
somewhat low $\chi^2$ minimum (see Table~\ref{tab:conf}), yet
compatible with statistical fluctuations.  For example, $\chi^2 =
49.2$ for 63~dof corresponds to a goodness-of-fit of 89.8\%.

\begin{figure} \centering
  \includegraphics[width=0.5\textwidth]{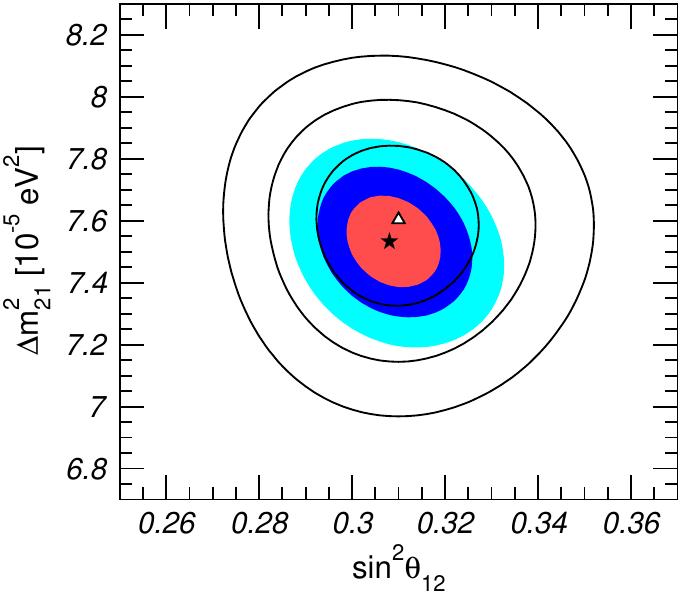}
  \caption{Impact of first JUNO data on the global determination of
    $\Dmq_{21}$ and $\sin^2 \theta_{12}$.  We show the $1\sigma$,
    $2\sigma$, and $3\sigma$ allowed regions (2 dof) without (black)
    and with (colored) JUNO data.}
  \label{fig:12_wo_w_JUNO}
\end{figure}

We include these results in the global NuFIT~6.1
analysis~\cite{nufit-6.1, Esteban:2024eli} by adding the corresponding
$\chi^2_\text{JUNO,cnf~2} (\Dmq_{21}, \theta_{12}, \theta_{13})$
marginalized over $\Dmq_{3\ell}$.  This way, no information on
$\Dmq_{3\ell}$ is included in the combination, so that the global
result is only based on information published by the JUNO
collaboration (including the systematic uncertainties relevant for the
analysis).  We refer the reader to Ref.~\cite{nufit-6.1} for the full
set of figures and oscillation parameter values for this global
analysis update, that also includes the latest results from
SNO+~\cite{SNO:2025chx} and
IceCube~\cite{IceCubeCollaboration:2023wtb}.  To highlight the impact
of first JUNO results on $\Dmq_{21}$ and $\sin^2 \theta_{12}$, we show
in Fig.~\ref{fig:12_wo_w_JUNO} the allowed regions on these parameters
without and with JUNO data.  As the figure shows, JUNO already
dominates the global analysis.  Numerically, we find the following
allowed parameter values within $1\sigma$
\begin{equation}
  \begin{aligned}
    \sin^2 \theta_{12}
    &= 0.3096_{-0.0073}^{+0.0057} \,,
    \qquad \theta_{12} = 33.81_{-0.46}^{+0.35} \,,
    \\[2mm]
    \Dmq_{21}
    &= (7.530_{-0.097}^{+0.096})\times 10^{-5}~\eVq \,,
  \end{aligned}
\end{equation}
in good agreement with the global combination in
Ref.~\cite{Capozzi:2025ovi}.


\section{Sensitivity to $|\Dmq_{3\ell}|$ and the Mass Ordering}
\label{sec:order}

We now delve into the information on $\Dmq_{3\ell}$ from the JUNO
analysis that we have so far marginalized over.  We stress again that
this goes beyond the analyses published by the JUNO collaboration, and
has therefore an exploratory nature.  The following results need to be
corroborated by official analyses including relevant systematics and
increased statistics.

\begin{figure} \centering
  \includegraphics[height=0.37\textwidth]{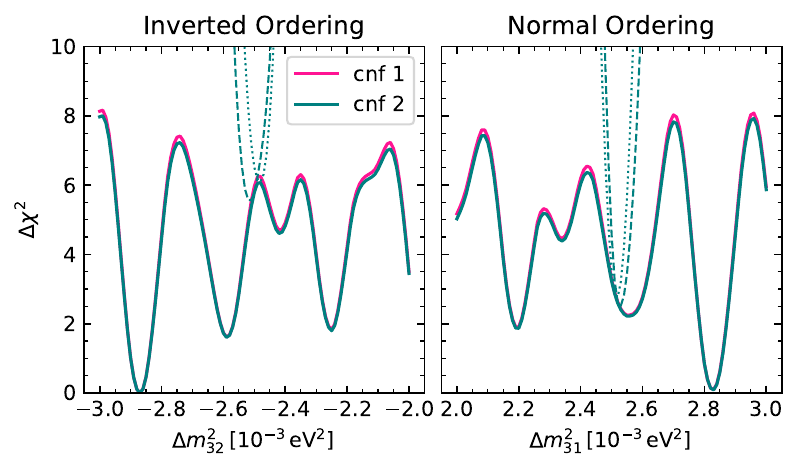} \hfill
  \includegraphics[height=0.345\textwidth]{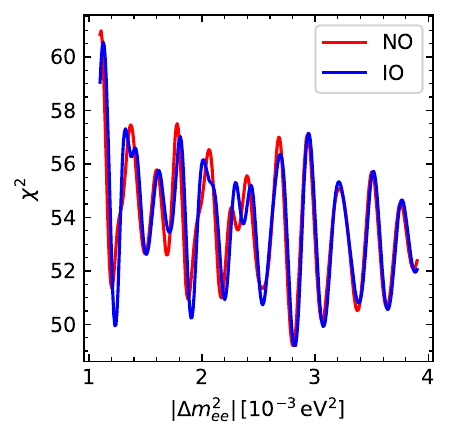}
  \caption{Dependence of $\chi^2_\text{JUNO}$ on $\Dmq_{3\ell}$ for
    cnf~1 and 2 (two left panels) and on $\Dmq_{ee}$ with an extended
    range for cnf~2 (right panel).  In the left panels, dashed and
    dotted curves correspond to $\Delta\chi^2_\text{JUNO}$ (cnf~2)
    combined with the determination of $|\Dmq_{3\ell}|$ from NuFIT-6.1
    global data without and with SK-ATM, respectively.}
  \label{fig:MO-1-2}
\end{figure}

The sensitivity to $\Dmq_{3\ell}$ emerges from ``fast'' oscillations
with an amplitude controlled by $\sin^22\theta_{13}$, see the second
line in Eq.~\eqref{eq:Pee}.  These small oscillation features are
clearly visible in the predicted spectra, for instance in
Figs.~\ref{fig:spectrum} and~\ref{fig:par12-1-2}.  In the left panel
of Fig.~\ref{fig:MO-1-2}, we show the one-dimensional projection
$\Delta\chi^2_\text{JUNO} (\Dmq_{3\ell})$ after marginalizing over
$\Dmq_{21}$ and $\theta_{12}$ for configurations cnf~1 and~2.  We have
verified that marginalizing over $\Dmq_{21}$ and $\theta_{12}$ or
fixing them to their best-fit values gives identical results for the
$\Dmq_{3\ell}$ determination.  In other words, correlations between
the `$12$' parameters and $\Dmq_{3\ell}$ in the current JUNO analysis
are negligible.  As mentioned above, we impose the global
determination of $\theta_{13}$, $\sin^2\theta_{13} = 0.022 \pm
0.00056$, as an external constraint on the fit.
The profiles shown in Fig.~\ref{fig:MO-1-2} indicate several preferred
values of $\Dmq_{3\ell}$, suggesting the presence of an oscillatory
feature in the data with a $\chi^2$ difference of $4 \lesssim \Delta
\chi^2 \lesssim 8$ among different assumed phases for the oscillation.

For JUNO alone, the $\chi^2$ difference between the best fits for NO
and IO is negligibly small, $|\Delta\chi^2_\mathrm{IO-NO}| < 10^{-3}$
for all configurations.  We therefore conclude that the current
dataset shows no sensitivity to the MO on its own.  The JUNO-only best
fits for the two orderings correspond to
\begin{equation}
  \Dmq_{31} =  2.83\times 10^{-3}~\eVq \,,\qquad
  \Dmq_{32} = -2.87\times 10^{-3}~\eVq \,.
\end{equation}
In the right panel of Fig.~\ref{fig:MO-1-2}, we show for cnf~2 an
extended range in $|\Dmq_{3\ell}|$, plotted in terms of the effective
squared-mass difference relevant for $\bar\nu_e$ disappearance in
reactors, $\Dmq_{ee} \equiv c_{12}^2 \Dmq_{31} + s_{12}^2
\Dmq_{32}$~\cite{Nunokawa:2005nx}.  The best-fit points for both
orderings correspond to the same value of $|\Dmq_{ee}| \approx
2.8\times 10^{-3}~\eVq$ within good accuracy.  However, as the figure
shows, there are multiple minima with $\Delta\chi^2 \lesssim 2$, and
therefore the particular location of the best-fit point is of no
significance.

However, global oscillation data constrain $|\Dmq_{3\ell}|$ with
percent-level precision.  The NuFIT-6.1 results are (in units of
$10^{-3}~\eVq$)
\begin{equation}
  \label{eq:dmq3l-BF}
  \begin{aligned}
    \Dmq_{31} &= 2.521_{-0.018}^{+0.026} \,,
    &\qquad \Dmq_{32} &= -2.510_{-0.023}^{+0.024}
    &\qquad &\text{(NuFIT-6.1 w/o SK-ATM)},
    \\
    \Dmq_{31} &= 2.511_{-0.020}^{+0.021} \,,
    &\qquad \Dmq_{32} &= -2.484 \pm 0.020 &
    \qquad &\text{(NuFIT-6.1 with SK-ATM)},
  \end{aligned}
\end{equation}
where we provide results for both mass orderings and for two variants
of the global analysis, with and without the Super-Kamiokande and
IceCube-24 atmospheric neutrino $\chi^2$ tables added to the global
analysis (here labeled ``w/o SK-ATM'' and ``with SK-ATM'' for
simplicity), see Refs.~\cite{Esteban:2024eli, nufit-6.1} for details.
We include this information into the JUNO analysis by adding an
external prior to $\chi^2_\text{JUNO}$ given by two-sided parabolas as
a function of $\Dmq_{3\ell}$ with their minimum set to zero for both
orderings.  In this way, we isolate the MO discrimination power due to
the combination of JUNO data and external information on
$|\Dmq_{3\ell}|$.

The results are shown as dashed and dotted curves in the left panels
of Fig.~\ref{fig:MO-1-2} for cnf~2 (for cnf~1, the results are very
similar).  We see that the agreement of JUNO with the remaining global
data is somewhat better for NO, implying a preference for NO over IO.
Assuming cnf~2, we find $\Delta\chi^2_\mathrm{IO-NO} = 3.05\, (3.28)$
for NuFIT without (with) SK-ATM, with similar results also for cnf~1
as seen in Table~\ref{tab:conf}.  We quantify this result in terms of
a hypothesis test and $p$-values in the next subsection.

The sensitivity of the dataset can be determined by the quantity
$T_0$~\cite{Blennow:2013oma}, obtained from the Asimov datasets
assuming the corresponding NuFIT-6.1 best fit values of $\Dmq_{3\ell}$
as ``true values''.  For both cnf~1 and~2 we find $|T_0|\approx 1.4$
(1.7) without (with) SK-ATM (see Table~\ref{tab:conf} for detailed
numbers).  As discussed in Ref.~\cite{Blennow:2013oma}, $|T_0|$ and
$\sqrt{|T_0|}$ roughly correspond to the expected value of
$|\Delta\chi^2_\mathrm{IO-NO}|$ and to the median MO sensitivity in
units of standard deviations, respectively.  This naively suggests
that JUNO data might be slightly more sensitive than expected.  We
quantify this in the next subsection.  For JUNO data alone (without
the external constraint on $|\Dmq_{3\ell}|$) the Asimov sensitivity to
the MO is negligible, $|T_0| \lesssim 10^{-3}$, hence current data on
its own shows no sensitivity to the MO.


\subsection{Monte Carlo simulation of MO result}
\label{sec:MC}

From the above results on the $T_0$ values, we expect a relatively
small --~but non-negligible~-- sensitivity to the MO from this data
combination.  In this subsection, we investigate by Monte Carlo
simulation the stability of the obtained preference for NO with
respect to statistical fluctuations, and perform a MO hypothesis test
to evaluate the $p$-value of IO.  To this aim, we simulate a large set
of randomly generated pseudo-data, and compare the result from true
data to the expected distribution.  For this study, we adopt
configuration~2.

The procedure is as follows.  First, we assume ``true'' oscillation
parameters: $\sin^2\theta_{12} = 0.31$, $\Dmq_{21} = 7.5\times
10^{-5}~\eVq$, $\sin^2\theta_{13} = 0.022$, and one of the four cases
for the $\Dmq_{3\ell}$ best fit points given in
Eq.~\eqref{eq:dmq3l-BF}.  We calculate the reactor signal in JUNO for
these parameters and add up the background prediction.  Then, we
generate random pseudo-data in each bin, Poisson-distributed around
the total prediction.  This pseudo-data set is analysed assuming both
NO and IO.  Since, as discussed above, correlations with
$\sin^2\theta_{12}$ and $\Dmq_{21}$ are negligible, we first minimize
with respect to these parameters, then add the NuFIT-6.1 constraints
on $|\Dmq_{3\ell}|$ as external prior as described above, and we
finally minimize with respect to $|\Dmq_{3\ell}|$ to compute
\begin{align}
  \Delta\chi^2_\mathrm{IO-NO}
  \equiv \chi^2_\text{min}(\text{IO}) - \chi^2_\text{min}(\text{NO}) \,.
\end{align}
This procedure is repeated $10^5$ times and each
$\Delta\chi^2_\mathrm{IO-NO}$ value is stored in a histogram.

\begin{figure}
  \centering \includegraphics[width=\textwidth]{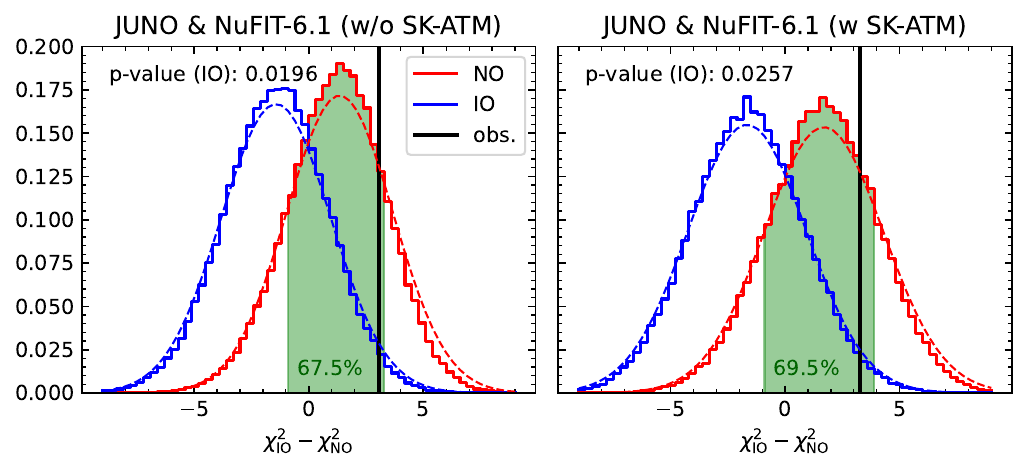}
  \caption{Monte Carlo simulation of the $\Delta
    \chi^2_\mathrm{IO-NO}$ distribution for JUNO combined with the
    NuFIT-6.1 constraint on $|\Dmq_{3\ell}|$ without SK-ATM (left
    panel) and with SK-ATM (right panel).  Black vertical lines
    indicate the values obtained by the observed data.  The green
    shaded part in the left (right) panel contains 67.5\% (69.5\%) of
    the histogram for NO.  Dashed curves show the Gaussian
    approximation based on the $T_0$-value from the Asimov data set.}
  \label{fig:MC}
\end{figure}

The corresponding distributions are shown in Fig.~\ref{fig:MC}.  The
dashed curves in the figure show a Gaussian distribution with mean
given by $T_0$ (see Table~\ref{tab:conf}) and standard deviation
$2\sqrt{|T_0|}$, which is the expected distribution of
$\Delta\chi^2_\mathrm{IO-NO}$~\cite{Qian:2012zn, Blennow:2013oma},
indicating reasonable agreement.  The $p$-value is obtained by
comparing the observed value with the MC distribution, leading to
$p$-values for IO --~defined as the probability to obtain a value for
$\Delta\chi^2_\mathrm{IO-NO}$ equal to or larger than the observed one
if IO were true~-- of 1.96\% (2.57\%) without (with) SK-ATM.  When
converted into Gaussian standard deviations, this corresponds to
$2.3\sigma$ ($2.2\sigma$), somewhat higher than the naive
$\sqrt{\Delta\chi^2_\mathrm{IO-NO}} \approx 1.75\sigma$ ($1.81\sigma$)
would suggest.

On top of that, as it is clear from the figure, the observed value of
$\Delta\chi^2_\mathrm{IO-NO}$ is not very unusual if the true ordering
is normal: the green shaded regions in Fig.~\ref{fig:MC} indicate
approximately the $\pm 1\sigma$ expected interval for
$\Delta\chi^2_\mathrm{IO-NO}$ assuming NO, and the observed value is
inside it.  Hence, although the observed value is larger than the mean
(\textit{i.e.}, larger than $T_0$), such an upward fluctuation would
not be very rare and is well within expectations.

We conclude this subsection with the following comment. Current JUNO
data on its own -- without external constraints on $\theta_{13}$ and
$\Dmq_{3\ell}$ -- cannot yet constrain these parameters with a
precision comparable to the global oscillation data. However, once the
external information is imposed, the percent-level precision from
global data on these parameters (up to the MO ambiguity) leads to a
very specific prediction for the oscillation pattern in JUNO, whose
comparison with data provides already a non-negligible sensitivity to
the MO. The Monte Carlo study presented here quantifies this, and
shows how likely (or unlikely) it is that the better match for NO than
for IO happens by pure chance.


\subsection{Robustness of the Mass Ordering Sensitivy}
\label{sec:robust}

In order to study the robustness of the MO sensitivity that we find
with our favoured configuration against possible unreported sources of
systematic uncertainties, we have performed a series of fits in which
some of the assumptions in cnf~2 are severely modified.  While we do
not necessarily suggest that these modifications are realistic, they
serve as illustrative examples how the MO sensitivity could
potentially be affected.

\begin{figure}
  \centering
  \includegraphics[width=0.54\textwidth]{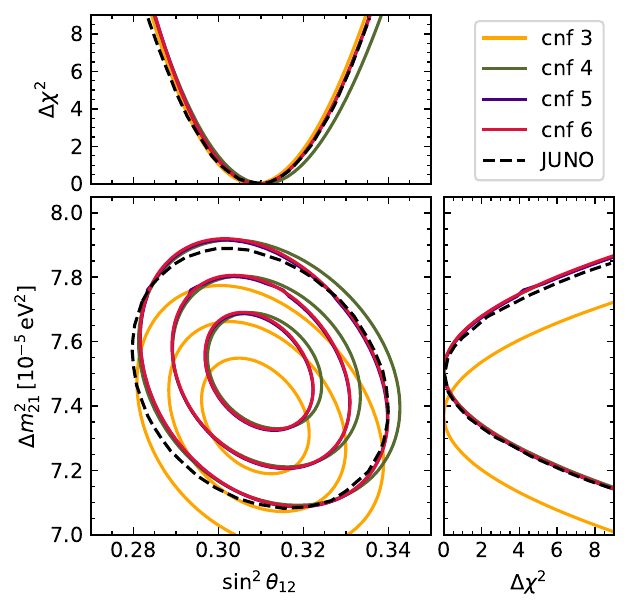}
  \caption{Determination of $\Dmq_{21}$ and $\sin^2\theta_{12}$ for
    cnf~3--6 (see Table~\ref{tab:conf}) as labeled in the figure
    compared to the JUNO results (black dashed line).  Regions are
    shown for $1\sigma$, $2\sigma$, $3\sigma$ (2~dof).  Results for
    cnf~5 and cnf~6 are nearly identical and curves overlap.}
  \label{fig:par12-3-6}
\end{figure}

\begin{figure}
  \includegraphics[width=0.6\textwidth]{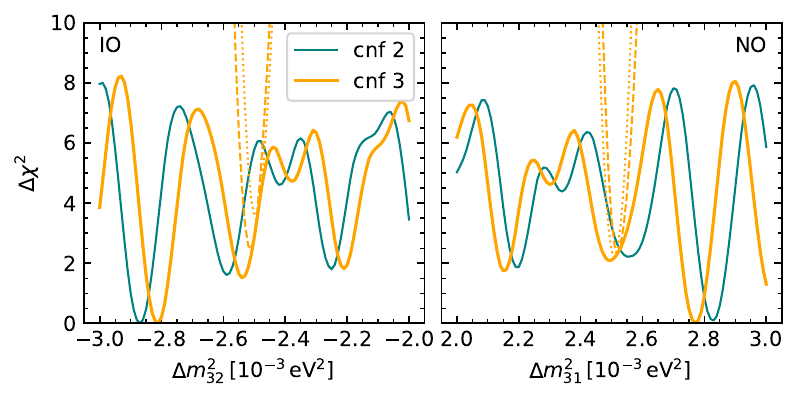}
  \includegraphics[width=0.4\textwidth]{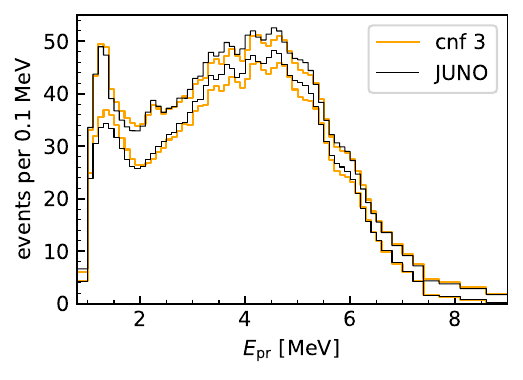}
  \\
  \includegraphics[width=0.6\textwidth]{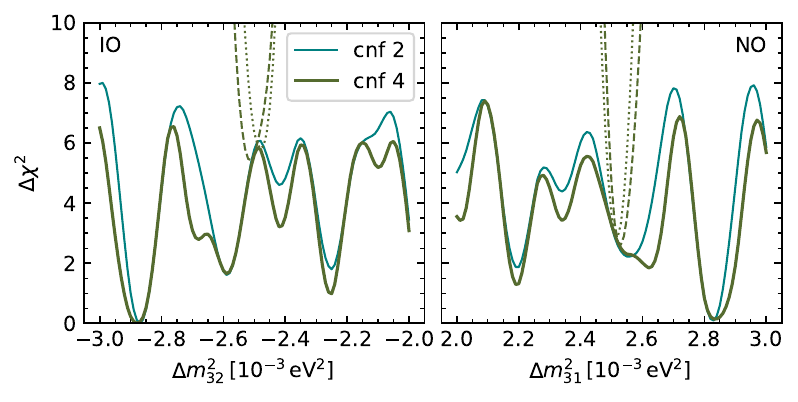}
  \includegraphics[width=0.4\textwidth]{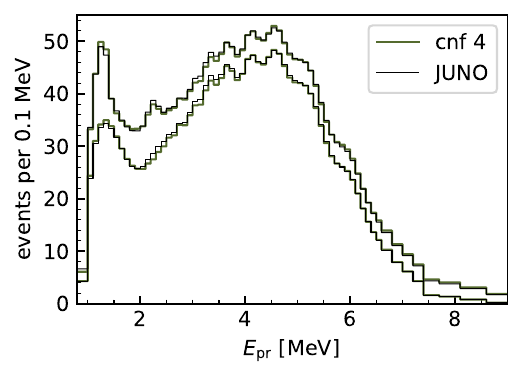}
  \\
  \includegraphics[width=0.6\textwidth]{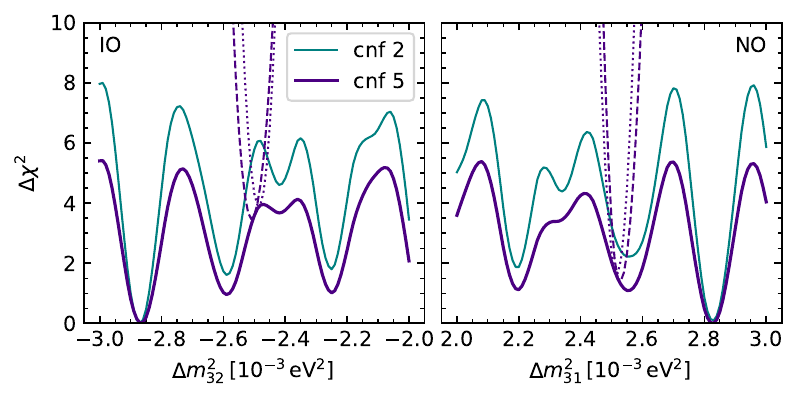}
  \includegraphics[width=0.4\textwidth]{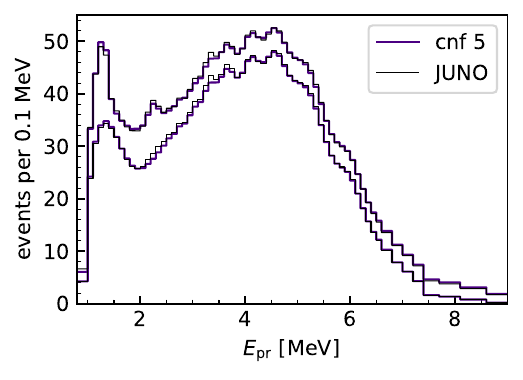}
  \\
  \includegraphics[width=0.6\textwidth]{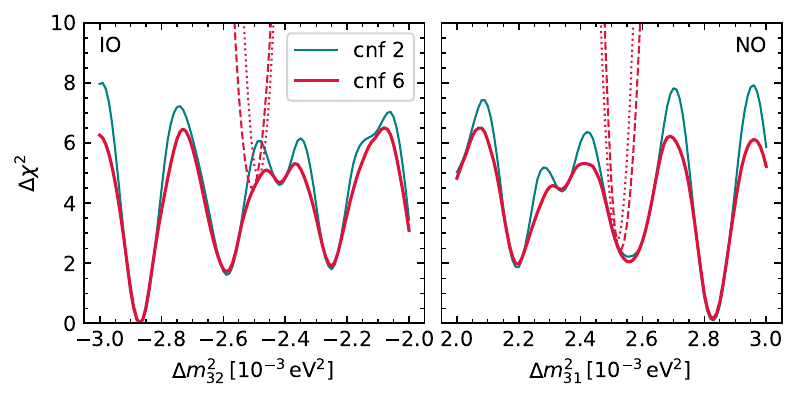}
  \includegraphics[width=0.4\textwidth]{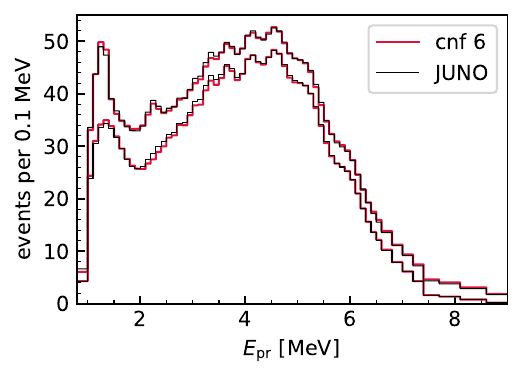}
  \caption{JUNO analysis variants cnf~3--6 (see Table~\ref{tab:conf}).
    Two left panels: dependence of $\Delta\chi^2_\text{JUNO}$ on
    $\Dmq_{3\ell}$ alone (solid) and in combination with NuFIT-6.1
    (dashed w/o SK-ATM, dotted with SK-ATM).  Right: Best-fit reactor
    neutrino spectra without pull shifts (lower histograms) and
    reactor neutrino+background spectra with pull shifts (higher
    histograms), compared to the official spectra from the JUNO
    collaboration (black).}
  \label{fig:MO-3-6}
\end{figure}

We summarize our findings with configurations cnf~3 to 6 listed in
Table~\ref{tab:conf}.  We focus on effects which are most likely to
affect the determination of $\Dmq_{3\ell}$.  In cnf~3 and~4 we study
the impact of changes in the systematic uncertainties affecting the
energy scale, while cnf~5 and~6 explore the impact of changes in the
systematic uncertainties affecting the energy resolution.  For each
case, we verify how the changes modify the determination of the ``12''
parameters and the MO.  The results are shown in
Figs.~\ref{fig:par12-3-6} and~\ref{fig:MO-3-6}.

The effect of a shift in the energy scale --~which we have
parametrized with the rescaling factor $r_\text{n.l.}$~-- is
illustrated with cnf~3, which is identical to cnf~2 except for an
energy scale shift by $+2.4\%$ (\textit{i.e.}, $r_\text{n.l.} =
1.024$).  We see in the upper panels in Fig.~\ref{fig:MO-3-6} that
this has a relatively strong impact on $\Delta\chi^2_\mathrm{IO-NO}$
(see also Table~\ref{tab:conf}). Note that this modification does not
affect the sensitivity to the $\Dmq_{3\ell}$-induced oscillations
itself (contrary to the modifications discussed in the following for
cnf~4,5,6); its impact emerges from a relative shift of the JUNO and
external constraints on $|\Dmq_{3\ell}|$, which may or may not lead to
a reduced value of $\Delta\chi^2_\mathrm{IO-NO}$. However, both
squared-mass splittings $\Dmq_{21}$ and $\Dmq_{3\ell}$ are affected in
the same way by such an energy rescaling, as can be seen in the
corresponding regions in Fig.~\ref{fig:par12-3-6}.  Furthermore, the
predicted spectrum no longer matches the official one provided by the
JUNO collaboration, as seen in the upper right panel in
Fig.~\ref{fig:MO-3-6}.  We conclude that this type of ``systematic
shift'' has a potentially strong impact on the MO, but if we take the
JUNO determination of $\Dmq_{21}$ as granted, such a shift is
essentially eliminated.  Quantitatively, the $+2.4\%$ shift chosen
here for illustration leads roughly to a $1\sigma$ shift in
$\Dmq_{21}$.  Hence, we would need a deviation of $\Dmq_{21}$ of order
$1\sigma$ in order for this to have a sizeable impact.

\begin{table}
  \centering\scriptsize
  \begin{tabular}{|l|ccccccccccc|}
    \hline
    $r_\text{n.l.}$
    & 0.975 & 0.980 & 0.985 & 0.990 & 0.995 & 1.000
    & 1.005 & 1.010 & 1.015 & 1.020 & 1.025
    \\
    $\# \sigma_\text{scl}$
    & $-5\sigma$ & $-4\sigma$ & $-3\sigma$ & $-2\sigma$ & $-1\sigma$ & $0$
    & $+1\sigma$ & $+2\sigma$ & $+3\sigma$ & $+4\sigma$  & $+5\sigma$
    \\
    \hline
    $\Delta\chi^2$ (w/o SK-ATM)
    & $\hphantom{+}1.01$ & 1.79 & 2.49 & 3.03 & 3.20 & 3.05
    & 2.61 & 2.05 & 1.47 & 0.83 & 0.26
    \\
    $\Delta\chi^2$ (w SK-ATM)
    & $-0.05$ & 0.72 & 1.49 & 2.24 & 2.87 & 3.28
    & 3.41 & 3.23 & 2.77 & 2.14 & 1.72
    \\
    \hline
  \end{tabular}
  \caption{$\Delta\chi^2_\mathrm{IO-NO}$ as a function of the energy
    re-scaling factor $r_\text{n.l.}$ from combining the
    $|\Dmq_{3\ell}|$ information from JUNO with the one from NuFIT-6.1
    without (third row) and with (fourth row) SK-ATM.  The second row
    shows the size of the energy shift in units of the standard
    deviation of the energy scale uncertainty.}
  \label{tab:rnl}
\end{table}

To further illustrate this effect, in Table~\ref{tab:rnl} we show
$\Delta\chi^2_\mathrm{IO-NO}$ as a function of the energy-scale shift
in steps of $\Delta r_\text{n.l.} = 0.5\%$.  This step corresponds to
one standard deviation of the energy scale uncertainty, \textit{i.e.},
cnf~3 is close to a $+5\sigma$ shift.  We see that a $1\sigma$ shift
of the energy scale has a relatively small impact, $\pm 2\sigma$
shifts can lead to a change in $\Delta\chi^2_\mathrm{IO-NO}$ of around
1 unit (corresponding to a change of 30\% in
$\Delta\chi^2_\mathrm{IO-NO}$), whereas shifts above $2\sigma$ may
imply a more dramatic change.  We note, however, that even in the $\pm
5\sigma$ range all $\Delta\chi^2_\mathrm{IO-NO}$ remain positive (with
one marginal exception), indicating that qualitatively the preference
for NO is robust with respect to this type of systematic, especially
taking into account that extensive calibration efforts of the JUNO
collaboration strongly constrain the energy scale~\cite{JUNO:2025fpc}.

If, instead of a fixed energy-scale shift, we consider an increased
energy-scale \emph{uncertainty}, the corresponding pulls will affect
the total spectrum (including the un-oscillated one).  Therefore, its
effects are limited by the measurement of the overall spectral shape.
Configuration 4 corresponds to such a case, introducing a
10-times-increased energy scale uncertainty $\sigma_\text{bias} =
5\%$.  Even with such a large energy scale uncertainty the solar
parameters in Fig.~\ref{fig:par12-3-6} can be matched well.  Although,
as seen in the second row of Fig.~\ref{fig:MO-3-6}, the $\Dmq_{3\ell}$
profile changes notably, the impact on the MO discrimination is small
(see the three lower rows in Table~\ref{tab:conf}).

We next move to configurations 5 and 6, which both feature a worsened
energy resolution.  This partially washes out the fast oscillations
and it therefore leads to a reduced significance of the oscillation
pattern, as clearly visible in the $\chi^2$ profiles in the two bottom
rows of Fig.~\ref{fig:MO-3-6}.  Configuration 5 illustrates the effect
of increasing the energy resolution by 30\%, \textit{i.e.}, including
a rescaling $r_\text{res} = 1.3$ in Eq.~\eqref{eq:rres}.  As seen in
Fig.~\ref{fig:par12-3-6}, the impact on the solar parameters and the
spectra is minimal, but the sensitivity to $\Dmq_{3\ell}$ is clearly
degraded as shown by the results in the third row in
Fig.~\ref{fig:MO-3-6} and quantified in Table~\ref{tab:conf}.  This is
in accordance with the well-known result that energy resolution is
crucial for this analysis.

Motivated by this, we finally investigate the impact of the
\emph{uncertainty} on the energy resolution, for which our default
choice is $\sigma_\text{res} = 5\%$.  We find that reducing this
uncertainty by a factor of 10 has a very small impact, leading only to
a negligible increase of $\Delta\chi^2_\mathrm{IO-NO}$.  Conversely,
by increasing the uncertainty we find rather small impact as long as
$\sigma_\text{res} \lesssim 20\%$.  Configuration~6 corresponds to an
extreme case where we have increased $\sigma_\text{res}$ to 40\%.  As
seen in Figs.~\ref{fig:par12-3-6} and~\ref{fig:MO-3-6}, even in this
case the impact on the ``12'' parameters and spectra is small, but
$\Delta\chi^2_\mathrm{IO-NO}$ is degraded.

We conclude from the results in this subsection that the systematics
considered here may affect the MO discrimination.  However, this is
only the case if they are pushed to (potentially unrealistically)
large values, which gives some confidence to the MO results.
Nevertheless, we stress that our selection of possible systematics is
certainly non-exhaustive.  For instance, one may imagine a non-linear
distortion of the energy scale affecting only part of spectrum.
Another, even more critical issue could be the presence of unknown,
percent-level bump-like features in the initial reactor anti-neutrino
spectra~\cite{Capozzi:2020cxm}.  This, however, will be addressed by
dedicated future reactor flux measurements by
JUNO-TAO~\cite{JUNO:2020ijm}, a near detector at $\sim 30$~m baseline
where oscillations are negligible, which will provide a
model-independent reference spectrum with sub-percent energy
resolution.


\section{Summary and implications for the global oscillation fit}
\label{sec:summary}

In this paper, we have presented an exploratory study on the
sensitivity of first data from the JUNO reactor experiment to the fast
oscillations induced by the ``atmospheric'' squared-mass splitting
$\Dmq_{3\ell}$.  We have carefully tuned our re-analysis of JUNO
spectral data to match as close as possible the published results,
which focus on the leading ``solar'' parameters $\sin^2\theta_{12}$
and $\Dmq_{21}$.  With this analysis at hand, we update the global
determination of these parameters by adding JUNO to the remaining
oscillation data, a combination released as
NuFIT-6.1~\cite{nufit-6.1}.  This analysis relies only on results
published by the JUNO collaboration and does not make use of any
information on $\Dmq_{3\ell}$ from JUNO.

Departing from that simulation, in the present paper we however go
beyond the results published by the JUNO collaboration, and study the
possible sensitivity to $\Dmq_{3\ell}$.  Our findings indicate the
presence of fast oscillations, favouring certain values of
$\Dmq_{3\ell}$ over others with a significance between $2\sigma$ and
$3\sigma$.  While JUNO data on its own has negligible sensitivity to
the neutrino MO, in combination with the independent determination of
$|\Dmq_{3\ell}|$ from world oscillation data we find that the global
value for NO is in somewhat better agreement with JUNO than for IO,
with $\Delta \chi^2_\mathrm{IO-NO} \approx 3$.

We have evaluated the statistical significance of this result by means
of a Monte Carlo simulation, finding that the $p$-value for IO is
around 2\% to 2.5\%, depending on the used external data.  Hence, we
obtain a $\sim 2\sigma$ preference of NO over IO.  While the obtained
result corresponds to a slight upward fluctuation compared to the
median sensitivity, the obtained value for $\Delta
\chi^2_\mathrm{IO-NO}$ is within the 68\% expected interval for NO and
hence well within the statistically expected range.

These results emerge purely from combining the information on
$|\Dmq_{3\ell}|$ from JUNO with the remaining oscillation data, and do
not take into account the information on the MO already available in
the global data without the JUNO contribution, which are $\Delta
\chi^2_\mathrm{IO-NO} = 1.49$ (5.91) without (with)
SK-ATM~\cite{nufit-6.1}. If we include this information now in the
combined analysis of JUNO and remaining world oscillation data, we
find an overall preference for NO with
\begin{equation}
  \begin{aligned}
    \Delta \chi^2_\mathrm{IO-NO} = 4.62
    & \qquad \text{(JUNO \& NuFIT-6.1 w/o SK-ATM)},
    \\
    \Delta \chi^2_\mathrm{IO-NO} = 9.41
    & \qquad \text{(JUNO \& NuFIT-6.1 with SK-ATM)},
  \end{aligned}
\end{equation}
where the two values correspond to either including or not including
the external $\chi^2$ tables from Super-Kamiokande and IceCube-24
atmospheric neutrino data.

Apart from this effect on the MO and the dramatic impact of JUNO data
on the ``12'' parameters discussed in Sec.~\ref{sec:solar}, the impact
on all other oscillation parameters is very small. In particular, for
$\Dmq_{3\ell}$ we find the following changes of the global fit results
when including the information from JUNO on this parameter (in units
of $10^{-3}~\eVq$):
\begin{equation}
  \begin{aligned}
    2.521_{-0.018}^{+0.026} &\:\to\: 2.529_{-0.021}^{+0.021} \,,
    &\enspace
    -2.500_{-0.023}^{+0.024} &\:\to\: -2.515_{-0.025}^{+0.031}
    &&\quad\text{(w/o SK-ATM),}
    \\[1mm]
    2.511_{-0.020}^{+0.021} &\:\to\: 2.519_{-0.020}^{+0.017} \,,
    &\enspace
    -2.483_{-0.020}^{+0.020} &\:\to\: -2.484_{-0.026}^{+0.026}
    &&\quad\text{(with SK-ATM).} 
  \end{aligned}
\end{equation}

We emphasize that the JUNO results concerning the MO need to be
considered preliminary.  They will have to be confirmed by dedicated
studies within the collaboration, including precise assessment and
publication of the relevant systematic uncertainties.  Because of
this, we have discussed a number of possible systematic uncertainties
and quantified their potential impact on the $|\Dmq_{3\ell}|$
determination.

In conclusion, the results shown here demonstrate the power of the
JUNO experimental setup, providing some intriguing sensitivity already
with its first 59.1 days of data.  These results make us look forward
to future data releases from this experiment.

\acknowledgments

We thank Anatael Cabrera for useful discussions.  This project is
funded by USA-NSF grant PHY-2210533 and by the European Union's
Horizon Europe research and innovation programme (Marie
Sk{\l}odowska-Curie Staff Exchange grant agreement
101086085-ASYMMETRY), and by ERDF ``A way of making Europe''.  It also
receives support from grants PID2022-\allowbreak 126224NB-\allowbreak
C21, PID2022-\allowbreak 142545NB-\allowbreak C21, PID2024-\allowbreak
156016NB-\allowbreak I00, PID2022-\allowbreak 136510NB-\allowbreak
C33, ``Unit of Excellence Maria de Maeztu'' award to the ICC-UB
CEX2024-001451-M, grant IFT ``Centro de Excelencia Severo Ochoa''
CEX2020-001007-S funded by MCIN/AEI/\allowbreak 10.13039/\allowbreak
501100011033, as well as from Basque Government IT1628-22 grant and
the UPV/EHU EHU-N25/11 grant.  This project is also supported by the
National Natural Science Foundation of China (12425506, 12375101,
12090060, and 12090064) and the SJTU Double First Class start-up fund
(WF220442604).

\section{Note added}

\begin{figure}
  \centering
  \includegraphics[width=0.4\textwidth]{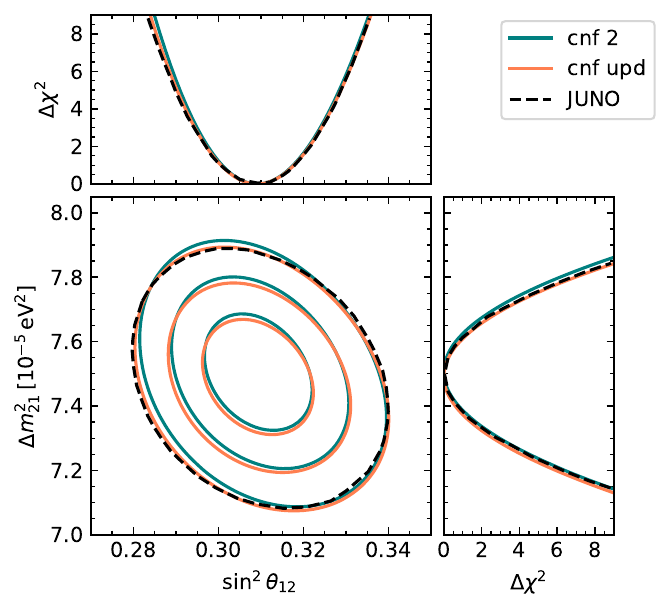}
  \includegraphics[width=0.55\textwidth]{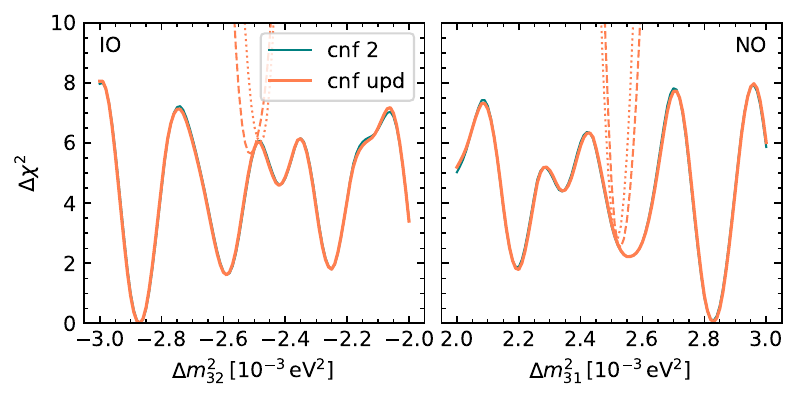}
  \caption{Left: Determination of $\Dmq_{21}$ and $\sin^2\theta_{12}$
    for the updated configuration compared to cnf~2 and to the JUNO
    results (black dashed line).  Regions are shown for $1\sigma$,
    $2\sigma$, $3\sigma$ (2~dof).  Right: $\Delta\chi^2$ as a function
    of $\Delta m^2_{3\ell}$ for the updated configuration compared to
    cnf~2 (curves nearly overlap).}
  \label{fig:update}
\end{figure}

After the acceptance of this work for publication a few minor technical improvements in the analysis code have been implemented. With these improvements, an excellent reproduction of the official JUNO results on $\theta_{12}$ and $\Delta m^2_{21}$ becomes possible without the 15\% rescaling of backgrounds as implemented for configuration~2 in the main text. The most relevant changes to the analysis are ($i$) using more accurate data on background spectra, ($ii$) adding the Fangchenggang power plant at a distance of 411.7~km with 12.1~GW$_{\rm th}$ to the reactor signal prediction, and ($iii$) adding an additional overall normalization uncertainty of 1.6\% on the selection efficiency from Tab.~1 of \cite{JUNO:2025gmd}, which gives a total $\sigma_{\rm norm} = 2.4\%$. With this new configuration no tuning of backgrounds or systematics with respect to the information provided in~\cite{JUNO:2025gmd} is necessary: fig.~\ref{fig:update} (left) shows that the updated analysis reproduces the official results on the solar parameters with excellent accuracy. The right panel shows that the $\Delta\chi^2$ profile as a function of $\Delta m^2_{31}$ is basically unaffected by the update, which implies that all results concerning the MO presented in the main text remain valid. 

\appendix

\section{Implementation of reactor neutrino fluxes}
\label{app:fluxreact}

In this appendix we give a detailed description of our implementation
of the reactor neutrino fluxes, based on the reconstructed spectra
provided by the Daya Bay collaboration.  We base our implementation on
the information available in the supplemental material of
Ref.~\cite{DayaBay:2025ngb}, which updates and supersedes the results
presented in Ref.~\cite{DayaBay:2021dqj} while following the same
approach and conventions.  The explicit construction of the reactor
fluxes can be divided into three steps.

First of all, in principle one should adapt the Daya-Bay flux
measurements --~which were collected with average fission fractions
$\bar{F}(\Nuc{235}{U}) : \bar{F}(\Nuc{238}{U}) :
\bar{F}(\Nuc{239}{Pu}) : \bar{F}(\Nuc{241}{Pu}) = 0.564 : 0.076 :
0.304 : 0.056$~-- to the specific isotopic composition of the JUNO
run.  However, lacking the detailed information on such composition,
we simply assume it to be the same as the Daya-Bay average, so that no
isotope correction is required and we can just extract the flux
information from the <<total>> part of the ancillary files of
Ref.~\cite{DayaBay:2025ngb}.  Concretely, the total $\bar{\nu}_e$
energy spectra weighted by IBD cross section $\Phi_i^0$ (with $i = 1
\dots 25$) is provided in the form of 25 energy-averaged bins (24
uniformely spaced between 1.8 and 7.8~MeV, plus a larger one from 7.8
to 9.5~MeV) together with the corresponding $25\times 25$ covariance
matrix $C_{ij}$ (which we extract from the complete $75\times 75$ one
given in the supplemental material, by discarding the 50 lines and
columns relative to the individual \Nuc{235}{U} and \Nuc{239}{Pu}
isotopes).  Thus $\Phi_i^0$ and $C_{ij}$ encode our knowledge of the
reactor flux and its uncertainties.

The second step consists in recasting the covariance matrix to the
pull formalism, so that it can be smoothly implemented into our codes
on the same footing as all the other systematic uncertainties.  To
this aim, we recall that $C_{ij}$ is a real symmetric
positive-definite matrix, hence it can be decomposed as $C_{ij} =
\sum_k O_{ik}\, O_{jk}\, D_k^2$ where $D_k^2$ are its positive
eigenvalues and $O_{ik}$ is a real orthogonal matrix.  Defining
$\Psi_{ik} \equiv O_{ik}\, D_k$ we can introduce the pull-dependent
$\bar{\nu}_e$ energy spectra $\Phi_i(\vec\xi_\text{flux}) = \Phi_i^0 +
\sum_k \Psi_{ik}\, \xi_\text{flux}^k$, where $\xi_\text{flux}^k$ (for
$k = 1 \dots 25$) are univariate zero-centered uncorrelated pulls.
Obviously $\Phi_i(0) = \Phi_i^0$, while the induced covariance matrix
is precisely $C_{ij} = \sum_k \Psi_{ik} \Psi_{jk}$ as required.

The final step involves converting the bin-averaged IBD-weighted
$\bar{\nu}_e$ spectra $\Phi_i(\vec\xi_\text{flux})$ into a continuous
flux shape $\phi(E_\nu, \vec\xi_\text{flux}) = \phi^0(E_\nu) + \sum_k
\psi_k(E_\nu)\, \xi_\text{flux}^k$ which can be inserted into
Eq.~\eqref{eq:signal} and properly integrated.  To achieve this, we
resort to cubic interpolation.  Concretely, we begin by writing:
\begin{equation}
  \label{eq:fluxcont}
  \phi^0(E_\nu) \equiv \phi_\text{huber}(E_\nu) \sum_n y_n^0\, \delta_n(E_\nu) \,,
  \qquad
  \psi_k(E_\nu) \equiv \phi_\text{huber}(E_\nu) \sum_n y_{nk}\, \delta_n(E_\nu)
\end{equation}
where $\delta_n(E_\nu)$ (for $n=1 \dots 25$) are piecewise cubic
polynomials used in the interpolation procedure, and
$\phi_\text{huber}(E_\nu)$ denotes the total reactor neutrino flux as
predicted in Ref.~\cite{Huber:2011wv} (calculated assuming the
Daya-Bay fission fractions quoted above).  Here
$\phi_\text{huber}(E_\nu)$ is introduced solely for the purpose of
normalization, to ensure that all the $y_n^0$ and $y_{nk}$ coefficient
(to be determined later) are of order 1.  Let us now recall the
definition of $\Phi_i^0$ (and, by extension, that of $\Psi_{ik}$):
\begin{equation}
  \Phi_i^0 = \big\langle \phi^0\, \sigma_\textsc{ibd} \big\rangle_i \,,
  \qquad
  \Psi_{ik} = \big\langle \psi_k\, \sigma_\textsc{ibd} \big\rangle_i \,,
  \qquad
  \big\langle f \big\rangle_i
  \equiv \frac{1}{E_{i+1} - E_i} \int_{E_i}^{E_{i+1}} f(E_\nu) \, dE_\nu
\end{equation}
where $\sigma_\textsc{ibd}(E_\nu)$ is the IBD cross-section.
Substituting Eq.~\eqref{eq:fluxcont} into these formulas, we obtain:
\begin{equation}
  \Phi_i^0 = \sum_n M_{in}\, y_n^0 \,,
  \qquad
  \Psi_{ik} = \sum_n M_{in}\, y_{nk} \,,
  \qquad
  M_{in} \equiv \big\langle \phi_\text{huber}\, \sigma_\textsc{ibd}\,
  \delta_n \big\rangle_i
\end{equation}
which can be immediately inverted, yielding:
\begin{equation}
  y_n^0 \equiv \sum_i M_{ni}^{-1}\, \Phi_i^0 \,,
  \qquad
  y_{nk} \equiv \sum_i M_{ni}^{-1}\, \Psi_{ik} \,.
\end{equation}
Substituting these expressions into Eq.~\eqref{eq:fluxcont} provides
the required interpolation.

\bibliographystyle{JHEPmod}
\bibliography{references}

\providecommand{\href}[2]{#2}\begingroup\raggedright\begin{thebibliography}{10}

\bibitem{JUNO:2025gmd}
{\scshape JUNO} collaboration, \emph{{First measurement of reactor neutrino
  oscillations at JUNO}},  \href{https://arxiv.org/abs/2511.14593}{{\ttfamily
  2511.14593}}.

\bibitem{JUNO:2022mxj}
{\scshape JUNO} collaboration, \emph{{Sub-percent precision measurement of
  neutrino oscillation parameters with JUNO}},
  \href{https://doi.org/10.1088/1674-1137/ac8bc9}{\emph{Chin. Phys. C}
  {\bfseries 46} (2022) 123001}
  [\href{https://arxiv.org/abs/2204.13249}{{\ttfamily 2204.13249}}].

\bibitem{JUNO:2015zny}
{\scshape JUNO} collaboration, \emph{{Neutrino Physics with JUNO}},
  \href{https://doi.org/10.1088/0954-3899/43/3/030401}{\emph{J. Phys. G}
  {\bfseries 43} (2016) 030401}
  [\href{https://arxiv.org/abs/1507.05613}{{\ttfamily 1507.05613}}].

\bibitem{Esteban:2024eli}
I.~Esteban, M.C.~Gonzalez-Garcia, M.~Maltoni, I.~Martinez-Soler, J.P.~Pinheiro
  and T.~Schwetz, \emph{{NuFit-6.0: updated global analysis of three-flavor
  neutrino oscillations}},
  \href{https://doi.org/10.1007/JHEP12(2024)216}{\emph{JHEP} {\bfseries 12}
  (2024) 216} [\href{https://arxiv.org/abs/2410.05380}{{\ttfamily
  2410.05380}}].

\bibitem{Petcov:2001sy}
S.T.~Petcov and M.~Piai, \emph{{The LMA MSW solution of the solar neutrino
  problem, inverted neutrino mass hierarchy and reactor neutrino experiments}},
  \href{https://doi.org/10.1016/S0370-2693(02)01591-5}{\emph{Phys. Lett. B}
  {\bfseries 533} (2002) 94}
  [\href{https://arxiv.org/abs/hep-ph/0112074}{{\ttfamily hep-ph/0112074}}].

\bibitem{Choubey:2003qx}
S.~Choubey, S.T.~Petcov and M.~Piai, \emph{{Precision neutrino oscillation
  physics with an intermediate baseline reactor neutrino experiment}},
  \href{https://doi.org/10.1103/PhysRevD.68.113006}{\emph{Phys. Rev. D}
  {\bfseries 68} (2003) 113006}
  [\href{https://arxiv.org/abs/hep-ph/0306017}{{\ttfamily hep-ph/0306017}}].

\bibitem{deGouvea:2005hk}
A.~de~Gouvea, J.~Jenkins and B.~Kayser, \emph{{Neutrino Mass Hierarchy, Vacuum
  Oscillations, and Vanishing $|U_{e3}|$}},
  \href{https://doi.org/10.1103/PhysRevD.71.113009}{\emph{Phys. Rev. D}
  {\bfseries 71} (2005) 113009}
  [\href{https://arxiv.org/abs/hep-ph/0503079}{{\ttfamily hep-ph/0503079}}].

\bibitem{Nunokawa:2005nx}
H.~Nunokawa, S.J.~Parke and R.~Zukanovich~Funchal, \emph{{Another Possible Way
  to Determine the Neutrino Mass Hierarchy}},
  \href{https://doi.org/10.1103/PhysRevD.72.013009}{\emph{Phys. Rev.}
  {\bfseries D72} (2005) 013009}
  [\href{https://arxiv.org/abs/hep-ph/0503283}{{\ttfamily hep-ph/0503283}}].

\bibitem{Minakata:2006gq}
H.~Minakata, H.~Nunokawa, S.J.~Parke and R.~Zukanovich~Funchal,
  \emph{{Determining Neutrino Mass Hierarchy by Precision Measurements in
  Electron and Muon Neutrino Disappearance Experiments}},
  \href{https://doi.org/10.1103/PhysRevD.74.053008}{\emph{Phys. Rev.}
  {\bfseries D74} (2006) 053008}
  [\href{https://arxiv.org/abs/hep-ph/0607284}{{\ttfamily hep-ph/0607284}}].

\bibitem{Blennow:2013vta}
M.~Blennow and T.~Schwetz, \emph{{Determination of the Neutrino Mass Ordering
  by Combining Pingu and Daya Bay II}},
  \href{https://doi.org/10.1007/JHEP09(2013)089}{\emph{JHEP} {\bfseries 09}
  (2013) 089} [\href{https://arxiv.org/abs/1306.3988}{{\ttfamily 1306.3988}}].

\bibitem{Cabrera:2020ksc}
A.~Cabrera et~al., \emph{{Synergies and prospects for early resolution of the
  neutrino mass ordering}},
  \href{https://doi.org/10.1038/s41598-022-09111-1}{\emph{Sci. Rep.} {\bfseries
  12} (2022) 5393} [\href{https://arxiv.org/abs/2008.11280}{{\ttfamily
  2008.11280}}].

\bibitem{Parke:2024xre}
S.J.~Parke and R.~Zukanovich-Funchal, \emph{{Mass ordering sum rule for the
  neutrino disappearance channels in T2K, NOvA, and JUNO}},
  \href{https://doi.org/10.1103/PhysRevD.111.013008}{\emph{Phys. Rev. D}
  {\bfseries 111} (2025) 013008}
  [\href{https://arxiv.org/abs/2404.08733}{{\ttfamily 2404.08733}}].

\bibitem{nufit-6.1}
I.~Esteban, M.C.~Gonzalez-Garcia, M.~Maltoni, I.~Martinez-Soler, J.P.~Pinheiro
  and T.~Schwetz, ``{NuFIT 6.1 (2026)}.'' \href{http://www.nu-fit.org}{\tt
  http://www.nu-fit.org}.

\bibitem{Maki:1962mu}
Z.~Maki, M.~Nakagawa and S.~Sakata, \emph{{Remarks on the unified model of
  elementary particles}}, \href{https://doi.org/10.1143/PTP.28.870}{\emph{Prog.
  Theor. Phys.} {\bfseries 28} (1962) 870}.

\bibitem{Kobayashi:1973fv}
M.~Kobayashi and T.~Maskawa, \emph{{CP Violation in the Renormalizable Theory
  of Weak Interaction}}, \href{https://doi.org/10.1143/PTP.49.652}{\emph{Prog.
  Theor. Phys.} {\bfseries 49} (1973) 652}.

\bibitem{Capozzi:2025wyn}
F.~Capozzi, W.~Giar{\`e}, E.~Lisi, A.~Marrone, A.~Melchiorri and A.~Palazzo,
  \emph{{Neutrino masses and mixing: Entering the era of subpercent
  precision}}, \href{https://doi.org/10.1103/PhysRevD.111.093006}{\emph{Phys.
  Rev. D} {\bfseries 111} (2025) 093006}
  [\href{https://arxiv.org/abs/2503.07752}{{\ttfamily 2503.07752}}].

\bibitem{deSalas:2020pgw}
P.F.~de~Salas, D.V.~Forero, S.~Gariazzo, P.~Mart{\'\i}nez-Mirav{\'e}, O.~Mena,
  C.A.~Ternes et~al., \emph{{2020 global reassessment of the neutrino
  oscillation picture}},
  \href{https://doi.org/10.1007/JHEP02(2021)071}{\emph{JHEP} {\bfseries 02}
  (2021) 071} [\href{https://arxiv.org/abs/2006.11237}{{\ttfamily
  2006.11237}}].

\bibitem{Vogel:1999zy}
P.~Vogel and J.F.~Beacom, \emph{{Angular Distribution of Neutron Inverse Beta
  Decay, $\bar\nu_e + p \to e^+ + n$}},
  \href{https://doi.org/10.1103/PhysRevD.60.053003}{\emph{Phys. Rev.}
  {\bfseries D60} (1999) 053003}
  [\href{https://arxiv.org/abs/hep-ph/9903554}{{\ttfamily hep-ph/9903554}}].

\bibitem{Strumia:2003zx}
A.~Strumia and F.~Vissani, \emph{{Precise quasielastic neutrino/nucleon
  cross-section}},
  \href{https://doi.org/10.1016/S0370-2693(03)00616-6}{\emph{Phys. Lett. B}
  {\bfseries 564} (2003) 42}
  [\href{https://arxiv.org/abs/astro-ph/0302055}{{\ttfamily
  astro-ph/0302055}}].

\bibitem{Tomalak:2025jtn}
O.~Tomalak, \emph{{Theory of inverse beta decay for reactor antineutrinos}},
  \href{https://arxiv.org/abs/2512.07956}{{\ttfamily 2512.07956}}.

\bibitem{Capozzi:2013psa}
F.~Capozzi, E.~Lisi and A.~Marrone, \emph{{Neutrino mass hierarchy and electron
  neutrino oscillation parameters with one hundred thousand reactor events}},
  \href{https://doi.org/10.1103/PhysRevD.89.013001}{\emph{Phys. Rev. D}
  {\bfseries 89} (2014) 013001}
  [\href{https://arxiv.org/abs/1309.1638}{{\ttfamily 1309.1638}}].

\bibitem{Li:2016txk}
Y.-F.~Li, Y.~Wang and Z.-z.~Xing, \emph{{Terrestrial matter effects on reactor
  antineutrino oscillations at JUNO or RENO-50: how small is small?}},
  \href{https://doi.org/10.1088/1674-1137/40/9/091001}{\emph{Chin. Phys. C}
  {\bfseries 40} (2016) 091001}
  [\href{https://arxiv.org/abs/1605.00900}{{\ttfamily 1605.00900}}].

\bibitem{Khan:2019doq}
A.N.~Khan, H.~Nunokawa and S.J.~Parke, \emph{{Why matter effects matter for
  JUNO}}, \href{https://doi.org/10.1016/j.physletb.2020.135354}{\emph{Phys.
  Lett. B} {\bfseries 803} (2020) 135354}
  [\href{https://arxiv.org/abs/1910.12900}{{\ttfamily 1910.12900}}].

\bibitem{Li:2025hye}
Y.-F.~Li, A.~Wang, Y.~Xu and J.-y.~Zhu, \emph{{Terrestrial Matter Effects on
  Reactor Antineutrino Oscillations: Constant vs. Fluctuated Density
  Profiles}},  \href{https://arxiv.org/abs/2511.15702}{{\ttfamily 2511.15702}}.

\bibitem{JUNO:2025fpc}
{\scshape JUNO} collaboration, \emph{{Initial performance results of the JUNO
  detector}},  \href{https://arxiv.org/abs/2511.14590}{{\ttfamily 2511.14590}}.

\bibitem{Ji:2019yca}
X.~Ji, W.~Gu, X.~Qian, H.~Wei and C.~Zhang, \emph{{Combined
  Neyman{\textendash}Pearson chi-square: An improved approximation to the
  Poisson-likelihood chi-square}},
  \href{https://doi.org/10.1016/j.nima.2020.163677}{\emph{Nucl. Instrum. Meth.
  A} {\bfseries 961} (2020) 163677}
  [\href{https://arxiv.org/abs/1903.07185}{{\ttfamily 1903.07185}}].

\bibitem{Capozzi:2025ovi}
F.~Capozzi, E.~Lisi, F.~Marcone, A.~Marrone and A.~Palazzo, \emph{{Updated
  bounds on the (1,2) neutrino oscillation parameters after first JUNO
  results}},  \href{https://arxiv.org/abs/2511.21650}{{\ttfamily 2511.21650}}.

\bibitem{SNO:2025chx}
{\scshape SNO+} collaboration, \emph{{Measurement of reactor antineutrino
  oscillations with 1.46 ktonne-years of data at SNO+}},
  \href{https://arxiv.org/abs/2511.11856}{{\ttfamily 2511.11856}}.

\bibitem{IceCubeCollaboration:2023wtb}
{\scshape IceCube} collaboration, \emph{{Measurement of atmospheric neutrino
  mixing with improved IceCube DeepCore calibration and data processing}},
  \href{https://doi.org/10.1103/PhysRevD.108.012014}{\emph{Phys. Rev. D}
  {\bfseries 108} (2023) 012014}
  [\href{https://arxiv.org/abs/2304.12236}{{\ttfamily 2304.12236}}].

\bibitem{Blennow:2013oma}
M.~Blennow, P.~Coloma, P.~Huber and T.~Schwetz, \emph{{Quantifying the
  sensitivity of oscillation experiments to the neutrino mass ordering}},
  \href{https://doi.org/10.1007/JHEP03(2014)028}{\emph{JHEP} {\bfseries 03}
  (2014) 028} [\href{https://arxiv.org/abs/1311.1822}{{\ttfamily 1311.1822}}].

\bibitem{Qian:2012zn}
X.~Qian, A.~Tan, W.~Wang, J.J.~Ling, R.D.~McKeown and C.~Zhang,
  \emph{{Statistical Evaluation of Experimental Determinations of Neutrino Mass
  Hierarchy}}, \href{https://doi.org/10.1103/PhysRevD.86.113011}{\emph{Phys.
  Rev. D} {\bfseries 86} (2012) 113011}
  [\href{https://arxiv.org/abs/1210.3651}{{\ttfamily 1210.3651}}].

\bibitem{Capozzi:2020cxm}
F.~Capozzi, E.~Lisi and A.~Marrone, \emph{{Mapping reactor neutrino spectra
  from TAO to JUNO}},
  \href{https://doi.org/10.1103/PhysRevD.102.056001}{\emph{Phys. Rev. D}
  {\bfseries 102} (2020) 056001}
  [\href{https://arxiv.org/abs/2006.01648}{{\ttfamily 2006.01648}}].

\bibitem{JUNO:2020ijm}
{\scshape JUNO} collaboration, \emph{{TAO Conceptual Design Report: A Precision
  Measurement of the Reactor Antineutrino Spectrum with Sub-percent Energy
  Resolution}},  \href{https://arxiv.org/abs/2005.08745}{{\ttfamily
  2005.08745}}.

\bibitem{DayaBay:2025ngb}
{\scshape Daya Bay} collaboration, \emph{{Comprehensive Measurement of the
  Reactor Antineutrino Spectrum and Flux at Daya Bay}},
  \href{https://doi.org/10.1103/PhysRevLett.134.201802}{\emph{Phys. Rev. Lett.}
  {\bfseries 134} (2025) 201802}
  [\href{https://arxiv.org/abs/2501.00746}{{\ttfamily 2501.00746}}].

\bibitem{DayaBay:2021dqj}
{\scshape Daya Bay} collaboration, \emph{{Antineutrino energy spectrum
  unfolding based on the Daya Bay measurement and its applications}},
  \href{https://doi.org/10.1088/1674-1137/abfc38}{\emph{Chin. Phys. C}
  {\bfseries 45} (2021) 073001}
  [\href{https://arxiv.org/abs/2102.04614}{{\ttfamily 2102.04614}}].

\bibitem{Huber:2011wv}
P.~Huber, \emph{{On the determination of anti-neutrino spectra from nuclear
  reactors}}, \href{https://doi.org/10.1103/PhysRevC.85.029901,
  10.1103/PhysRevC.84.024617}{\emph{Phys.Rev.} {\bfseries C84} (2011) 024617}
  [\href{https://arxiv.org/abs/1106.0687}{{\ttfamily 1106.0687}}].

\end{thebibliography}\endgroup

\end{document}